\documentclass[prl,twocolumn,superscriptaddress,showpacs,showkeys,preprintnumbers,amsmath,amssymb]{revtex4}
\usepackage{graphicx}
\usepackage{dcolumn}
\usepackage{bm}
\usepackage[active]{srcltx}
\usepackage{setspace}
\begin{document}

\title{Quantitative magnetic force microscopy on permalloy dots using an iron filled carbon nanotube probe} 

\author{ F.~Wolny }
\affiliation{Leibniz Institute for Solid State and Materials Research (IFW), Dresden, Helmholtzstrasse 20, 01069 Dresden, Germany}
\author{ Y.~Obukhov }
\affiliation{Department of Physics, The Ohio State University, Columbus, OH, 43210, USA}
\author{ T.~M\"uhl }
\author{ U.~Weissker }
\author{ S.~Philippi }
\author{ A.~Leonhardt }
\affiliation{Leibniz Institute for Solid State and Materials Research (IFW), Dresden, Helmholtzstrasse 20, 01069 Dresden, Germany}
\author{ P.~Banerjee}
\author{ A.~Reed}
\author{ G.~Xiang }
\author{ R.~Adur }
\author{ I.~Lee }
\author{ A.J~Hauser }
\author{ F.Y.~Yang }
\author{ D.V.~Pelekhov }
\affiliation{Department of Physics, The Ohio State University, Columbus, OH, 43210, USA}
\author{ B.~B\"uchner }
\affiliation{Leibniz Institute for Solid State and Materials Research (IFW), Dresden, Helmholtzstrasse 20, 01069 Dresden, Germany}
\author{ P.C.~Hammel }\email{hammel@mps.ohio-state.edu}
\affiliation{Department of Physics, The Ohio State University,
Columbus, OH, 43210, USA}

\date{\today}

\begin{abstract}
We have characterized a new Magnetic Force Microscopy (MFM) probe based on an iron filled carbon nanotube (FeCNT) using MFM imaging on permalloy (Py) disks saturated in a high magnetic field perpendicular to the disk plane. The experimental data is accurately modeled by describing the FeCNT probe as having a single magnetic monopole at its tip whose effective magnetic charge is determined by the diameter of the iron wire enclosed in the carbon nanotube and its saturation magnetization $4 \pi M_s \approx 2.2 \cdot10^4$ G. A magnetic monopole probe enables quantitative measurements of the magnetic field gradient close to the sample surface. The lateral resolution is defined by the diameter of the iron wire $\sim\!15$ nm and the probe-sample separation. As a demonstration, the magnetic field gradients close to the surface of a Py dot in domain and vortex states were imaged.
\end{abstract}

\maketitle
Magnetic Force Microscopy (MFM) is an essential technique for studying magnetic micro- and nanostructures \cite{MagDomains, Guntherodt:FirstMFM:jap:1987, Wiesendanger:magnetite:prb:2005, Chang:MagPartice:jap:1993, Hartmann:advances:2003}. It provides excellent lateral resolution  ($\sim 10$ nm) and is very sensitive to stray fields resulting from inhomogeneous sample magnetization (such as domains, magnetic vortices or variations in sample properties) or from boundaries of magnetic structures. MFM can provide quantitative information on magnetic structures and their magnetization reversal \cite{Hug:QuantativeMFM:apl:1998, Lohau:QuantitativeMFM:apl:2001, Hug:ExchBias:2005}, but this requires accurate probe characterization \cite{Hug:TipModel:jap:1992, Hartman:Tip:1994, Lohau:CoilMFM:jap:1999}.

An iron filled carbon nanotube (FeCNT) consists of an iron nanowire tens of nanometers in diameter and microns in length surrounded by a carbon shell \cite{Leonhardt2006}. FeCNTs are promising candidates as high resolution magnetic probes. A probe consisting of a single crystalline, single magnetic domain iron wire, with its high shape anisotropy and well known magnetization simplifies the interpretation of MFM data. Furthermore, the tough carbon shell protects the wire from mechanical damage and oxidation, making it a robust and long-lived probe. The first MFM images with FeCNT probes obtained on magnetic recording media demonstrated a high lateral magnetic resolution with magnetic feature sizes $\sim 20$--30\,nm \cite{Wolny2008, Wolny2010, winkler2006}.

Here we characterize the magnetic properties of a FeCNT MFM probe in detail. We show that the FeCNT can be modeled as a magnetic monopole with its magnetic charge determined by the iron wire diameter and its saturation magnetization $4\pi M^{\rm Fe}_s \approx 2.2 \cdot10^4$ G, typical for single crystal iron. We find this description to be valid for a wide range of probe sample separations, allowing us to use our monopole model to determine the carbon shell thickness at the end of the nanotube. The applicability of this simple and reliable monopole model makes the FeCNT probe an excellent tool for quantitative MFM measurements.
\begin{figure}
\includegraphics[width=1.0\columnwidth]{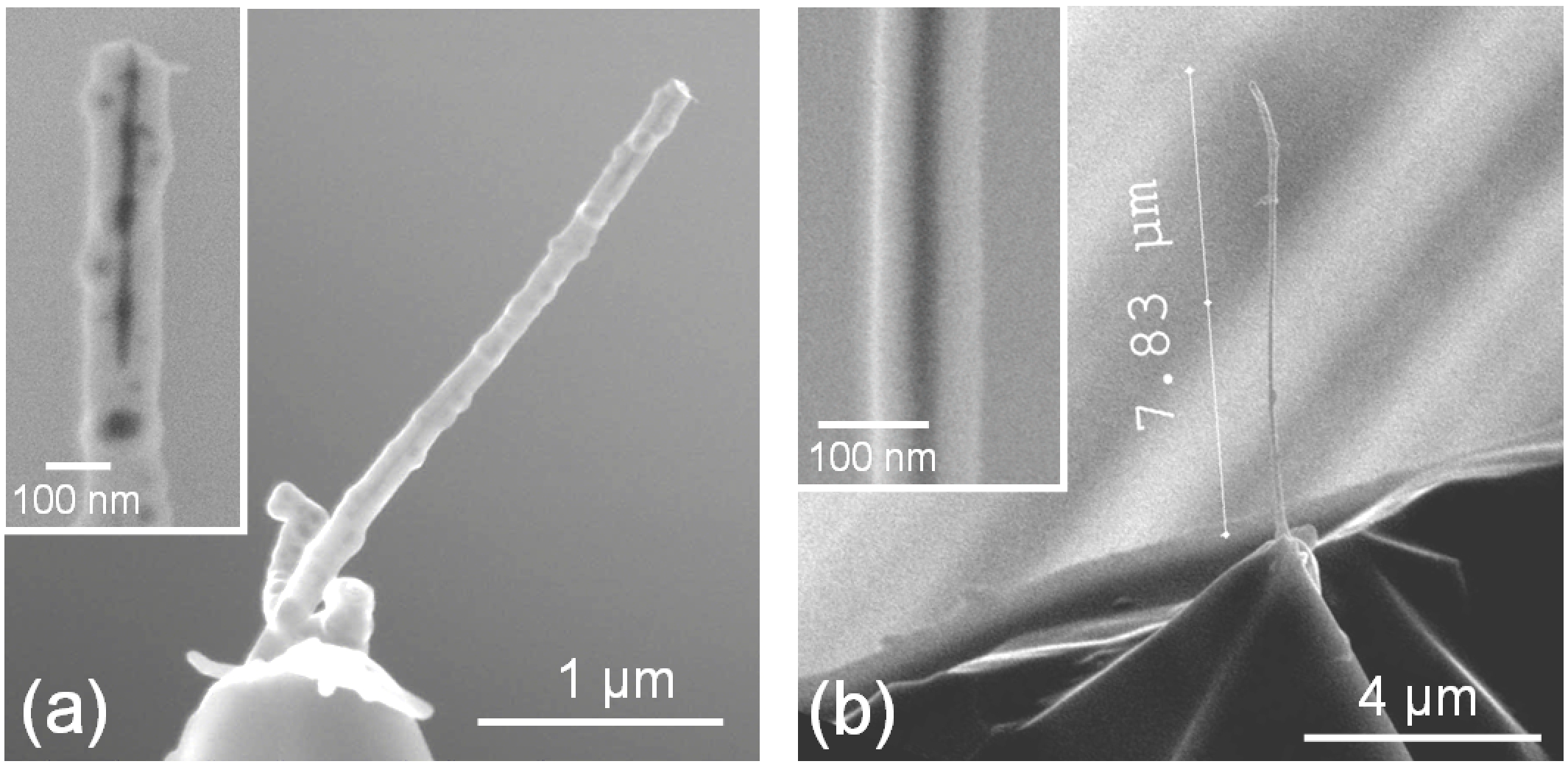}
\caption{SEM images of the two FeCNT MFM probes a) J2 and b) K2. The FeCNT is attached to the tip of a silicon cantilever. The insets show close-ups of the FeCNTs; the iron filling is darker than the carbon shells.}
\label{fig:J2_and_K2}
\end{figure}

The FeCNTs were prepared by chemical vapor deposition on catalyst coated silicon substrates with ferrocene as a precursor \cite{Mueller2006}. This method yields aligned multiwalled FeCNTs with a moderate distribution of lengths, diameters and filling degrees. The average FeCNT is $\approx 15\,\mu$m long and has a diameter of $\approx 100$ nm. It contains one long or several short iron wires with diameters ranging from 10 to 40\,nm. The MFM probes were prepared by attaching a single FeCNT onto a conventional tapping mode AFM cantilever with the help of a micromanipulator in a scanning electron microscope (SEM) \cite{Wolny2008}. Fig.~\ref{fig:J2_and_K2} shows SEM images of the FeCNT probes used in this experiment. A $\sim 3\,\mu$m long piece of FeCNT was attached to Probe J2 (Fig.~\ref{fig:J2_and_K2}a). As shown in the inset, the iron wire at its end is $\sim$ 500 nm long, and $\sim 10$--20 nm in diameter. This image was obtained by superimposing an SEM in-lens detector image with the inverted backscattered-electron-detector image of the same area. The iron filled parts are visible as dark regions. However, due to limited resolution the diameter of the iron filling is hard to evaluate precisely from SEM measurements alone. The FeCNT on probe K2 (Fig.~\ref{fig:J2_and_K2}b) is approximately 8 $\mu$m long, and is completely filled with an iron wire with a constant diameter of $\sim$ 20 nm (the inset shows a region in the middle of the FeCNT). This FeCNT was cut to the displayed length with a Focused Ion Beam (FIB) to remove unwanted iron particles from its end.  Both silicon cantilevers have a resonance frequency $f_0 \sim 13$ kHz and a spring constant $k \sim 130$ dyn/cm (0.13 N/m) measured using a hydrodynamic model \cite{sader:SpringConstant:64}.

\begin{figure}[b]
\includegraphics[width=1.0\columnwidth]{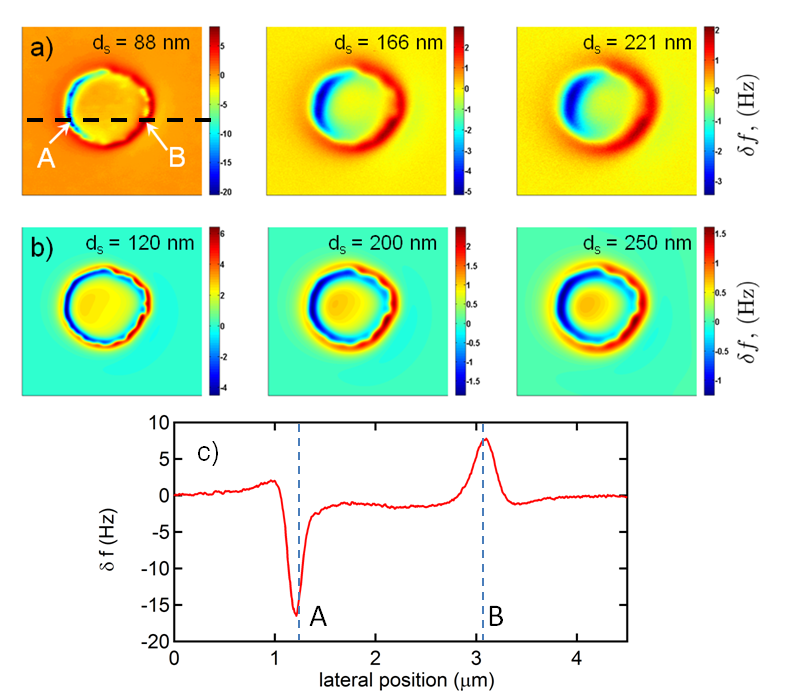}
\caption{a) MFM images of a Py disk in a $\sim 2$ T magnetic field perpendicular to the sample plane made with probe J2 at various probe-sample separations $d_s$; the scan area is $4.5 \times 4.5 \, \mu \rm m^2$. The $x$ and $y$ axes are horizontal and vertical, respectively. The points A and B show the position of the minimum and maximum cantilever frequency shift (see text). b) Simulations of the MFM images using the model shown in Fig.~\ref{fig:Model}. The simulations were performed for the same separations $d_s$ as in a), but a carbon shell thickness of 30 nm, determined as described below, was added. c) Line cut indicated by the black dashed line in panel a).}
\label{fig:MFMvsZ}
\end{figure}

The FeCNT probes were characterized by performing MFM on permalloy (Py) disks in a high magnetic field of $\sim 2$ T perpendicular to the Py film plane. The Py disk array was fabricated by photolithography and a lift-off process. The Py was deposited onto the silicon substrate with titanium adhesion and capping layers. The Py thickness is 40 nm, the disk diameter is 2.2 $\mu$m and the center-to-center spacing of the disks is 6 $\mu$m. The saturation magnetization of Py was measured by conventional Ferromagnetic Resonance (FMR) to be $4\pi M^{\rm Py}_s = 9530$ G. The MFM experiments were performed in high vacuum and at a temperature of 5 K. The cantilever deflection is detected by fiber-optic interferometry, and its frequency is monitored by a digital signal processor \cite{h:oboukhov.dsp}.

Typical MFM images measured with the FeCNT probe J2 at different probe-sample separations $d_s$, and at a temperature of 5 K are presented in Fig.~\ref{fig:MFMvsZ}a. The MFM signal is the cantilever frequency shift monitored during the scan at constant $d_s$ without SPM feedback. Force-distance curves (DC force on the cantilever as a function of probe height) were used to determine the probe touch point ($d_s=0$) with an accuracy of 10--15 nm. The microscope does not have vibration isolation inside the vacuum can, so the accuracy of the $d_s$ measurement is limited by motion of the probe in response to mechanical vibrations induced by boiling liquid helium. The DC deflection of the cantilever due to DC MFM forces during a scan was estimated to be less then 1 nm. The peak-to-peak cantilever oscillation was kept much smaller than $d_s$ and was usually set to 10 nm; for $d_s < 80$ nm it was reduced to 5 nm. The cantilever frequency shift $\delta \! f$ due to a force gradient is given by:
\[
  \delta \! f(x,y)=\frac{f_0}{2k}\frac{\partial F}{\partial z}(x,y)
\]
where $f_0$ is the cantilever resonance frequency, $k$ its spring constant and $\frac{\partial F}{\partial z}$ is the force gradient in the direction of the cantilever oscillation $\hat z$.

\begin{figure}[t]
\center{\includegraphics[width=0.65\columnwidth]{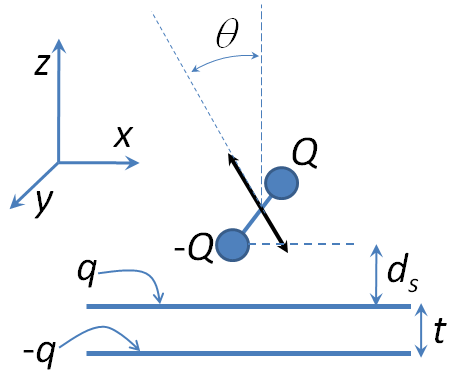}}
\caption{Monopole model of the FeCNT. A uniformly magnetized iron wire which is located at a distance $d_s$ from the sample surface can be defined by its two monopoles $Q$ and -$Q$. The sample, a Py disk of thickness $t$, is magnetized perpendicular to the disk plane in high external magnetic fields. The magnetization of the disk can be described by two sheets of magnetic charges with a charge density $q$ defined by the Py saturation magnetization. The black arrow indicates the direction of cantilever oscillation and the angle $\theta$ characterizes the cantilever tilt.}
\label{fig:Model}
\end{figure}

We calculate the MFM force gradient $\frac{\partial F}{\partial z}$ under two assumptions (see Fig.~\ref{fig:Model}). First, we consider the iron wire in the CNT to be uniformly magnetized along its long axis. In this case its magnetization can be described by two monopoles $Q$ and $-Q$ positioned at the ends of the wire. The monopole charge $Q = \pi d^2 M^{\rm Fe}_s / 4 $ is determined by the diameter $d$ of the iron wire and its saturation magnetization $M^{\rm Fe}_s$. Our experiments were performed in a magnetic field of 2 T, close to $4\pi M^{\rm Fe}_s$. If the iron wire is not parallel to the external field its magnetization will tilt slightly away from the wire axis reducing the monopole at the wire end. However, for moderate FeCNT tilt angles ($\leq 20$--30$^\circ$) the monopole description is still reasonable. We also assume the Py film to be saturated in the direction of the external field, perpendicular to the film plane. Consequently the magnetization of the Py film can be represented by two effective monopole layers with an effective charge per unit area $q = M^{\rm Py}_s$ defined by the saturation magnetization of Py (see Fig.~\ref{fig:Model}).

Since the stray field of the film falls off rapidly compared to the iron wire length, the influence of $Q$ associated with the remote end of the wire (see Fig.~\ref{fig:Model}) can be neglected. We can express the MFM force gradient induced by the sample's upper monopole layer as
\begin{eqnarray}
   \frac{\partial F}{\partial z}(x,y)
   & = &\int \left.
     \frac{\partial H_z} {\partial z} \right|_z (x-x',y-y')  \; \;  q(x',y')dx'dy'\nonumber\\
   & \equiv & \left. \frac{\partial H_z}{\partial z}\right|_z \ast q\nonumber
\end{eqnarray}
where ${\bf H}=-Q{\bf r}/r^3$ is the magnetic field created by the tip monopole $-Q$, $\bf r$ is the radius vector, and $\ast$ indicates a convolution. The total MFM force gradient created by both Py monopole layers can be written as
\[
  \frac{\partial F}{\partial z}
  = \left.\frac{\partial H_z}{\partial z}\right|_z\ast q
   -\left.\frac{\partial H_z}{\partial z}\right|_{z+t}\ast q
\]
where $t$ is the Py film thickness. This predicts the MFM image will be cylindrically symmetric about the center of the Py disk but we observe (see Fig.~\ref{fig:MFMvsZ}a) considerable asymmetry in the $\hat x$ (horizontal) direction. This can be explained by the tilt of the cantilever and its oscillation relative to the $z$ axis (Fig.~\ref{fig:Model}). In our setup this tilt is in the $xz$ plane and the tilt angle is $\theta = 15$--20$ ^\circ$. In this case the MFM force gradient is given by:
 \begin{equation}
   \frac{\partial \bf F}{\partial \bf l}
   =\left.\frac{\partial {\bf H}}{\partial \bf l}\right|_z\ast q
   -\left.\frac{\partial {\bf H}}{\partial \bf l}\right|_{z+t}\ast q
 \label{eq:ForceGradient}
 \end{equation}
where $\bf l$ is a vector parallel to the direction of cantilever oscillation. A calculation of $\frac{\partial \bf H}{\partial \bf l}$ in our geometry gives:
\begin{eqnarray}
  \frac{\partial {\bf H}}{\partial {\bf l}}&=&
  \frac{\partial H_x}{\partial x}\sin^2{\theta}
  +
  \left(\frac{\partial H_x}{\partial z}
  +     \frac{\partial H_z}{\partial x}\right)
  \sin{\theta}\cos{\theta}\nonumber\\
  &+&
  \frac{\partial H_z}{\partial z}\cos^2{\theta} \label{eq:FieldGradient}\\
  \frac{\partial H_x}{\partial x}
    &=&-Q\frac{r^2-3x^2}{r^5}\nonumber\\
  \frac{\partial H_x}{\partial z}
    &=&\frac{\partial H_z}{\partial x}=Q\frac{3xz}{r^5}\nonumber\\
  \frac{\partial H_z}{\partial z}
    &=&-Q\frac{r^2-3z^2}{r^5}\nonumber
\end{eqnarray}

Using Eq.~\ref{eq:ForceGradient}, we modeled MFM images of a Py disk. The FeCNT monopole and carbon shell thickness were adjusted to obtain the best agreement between experimental and simulated data. The shape of the disk is not exactly circular, therefore we generated the shape of the disk boundary according to the MFM image in Fig.~\ref{fig:MFMvsZ}a at $d_s = 88$\,nm. The results of this model are presented in Fig.~\ref{fig:MFMvsZ}b for the same values of $d_s$ as in the experimental data plus a carbon shell thickness of $\sim 30$\,nm (evaluated later) which increases the distance of the probe monopole to the sample surface. The parameters used in the model are: saturation magnetization of the iron wire in the FeCNT $4\pi M^{\rm Fe}_s = 2.2 \cdot 10 ^4$\,G, diameter of the iron wire $d=  16$ nm, which implies that the FeCNT monopole $Q = 3.5 \cdot 10 ^{-9}$ emu/cm ($3.5 \cdot 10 ^{-10}$\,Am), saturation magnetization of the Py film $4\pi M^{\rm Py}_s = 9530$\,G (measured independently by FMR) and cantilever tilt $\theta = 20 ^\circ$.

The simulations (Fig.~\ref{fig:MFMvsZ}b) given by the proposed model show good qualitative and quantitative agreement with the experimental data in Fig.~\ref{fig:MFMvsZ}a for large separations, but there is a considerable discrepancy at $d_s = 88$\,nm. In the experiment, the negative cantilever frequency shift on the left side of the Py disk (see point A in Fig.~\ref{fig:MFMvsZ}a, $d_s = 88$\,nm and corresponding line cut in panel c) is substantially bigger than that expected from the model. We attribute this to the electrostatic attraction between the FeCNT probe and the Py disk which adds a negative frequency shift. This effect becomes smaller at larger probe sample distances. For a quantitative comparison of the model with the experiment we choose the maximum positive frequency shift on the right side of the disk (point B in the image in Fig.~\ref{fig:MFMvsZ}a, $d_s = 88$\,nm). Point B is located outside of the Py disk boundary, so the probe-sample distance is bigger and the contribution of the electrostatic attraction to the total force is smaller. However, at small probe-sample distances the measurements will still have an error induced by electrostatic forces (the magnetic field gradient of the probe or sample at point B will thus be somewhat underestimated close to the sample surface).
\begin{figure}
\includegraphics[width=1.0\columnwidth]{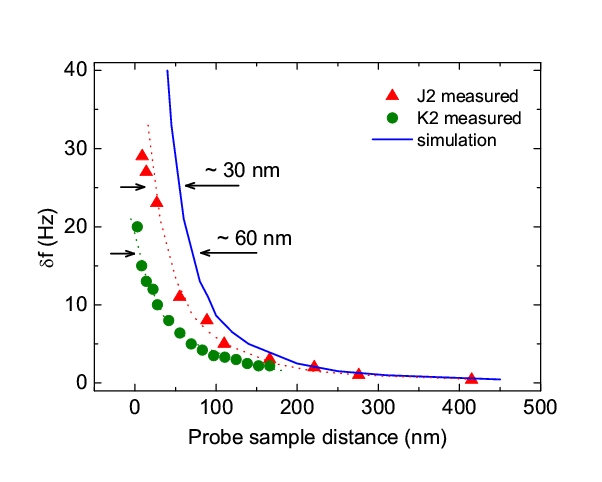}
\caption{MFM response at point B (Fig.~\ref{fig:MFMvsZ}a) for different probe sample distances $d_s$. The experimental data for the FeCNT probe J2 is displayed by triangles, the data for probe K2 by circles. The solid blue line is the simulation obtained using the monopole model in Fig.~\ref{fig:Model}. The experimental data and the model differ by a shift in $d_s$ (the dotted lines correspond to the shifted simulated curve). This shift corresponds to the distance between the actual position of the magnetic monopole and the measured $d_s$ touch point which is mainly determined by the FeCNT's carbon shell thickness.}
\label{fig:FreqVsZ}
\end{figure}

The $d_s$ dependencies of the frequency shifts for both FeCNT probes J2 and K2 are shown in Fig.~\ref{fig:FreqVsZ}. The measurements of the two probes are in very good quantitative agreement with the model (solid line, using Eq.~\ref{eq:ForceGradient} at point B; best agreement was obtained for $Q = 3.5\cdot 10^{-9}$\,emu/cm) once a shift in $d_s$ between the experimental data and the model reflecting the additional carbon shell thickness is included. The carbon wall increases the separation of the magnetic monopole from the tube end, and this thickness differs for various FeCNTs. For J2 we obtain a shell thickness $\sim 30$ nm which agrees with SEM images (inset to Fig.~\ref{fig:J2_and_K2}a). For K2 we find a thicker shell  $\sim 60$\,nm, possibly due to the roughness of the carbon shell. Since this FeCNT was FIB milled, the larger shift in $d_s$ may also reflect FIB induced damage to the end of the iron core. To show the agreement between experiment and model, we show the simulated curve after shifting it by the corresponding shell thickness to match the measured data (dotted lines in Fig.~\ref{fig:FreqVsZ}).

\begin{figure}
\hspace{-0.4cm}
\begin{minipage}[b]{4.2cm}
\includegraphics[width=4.6cm]{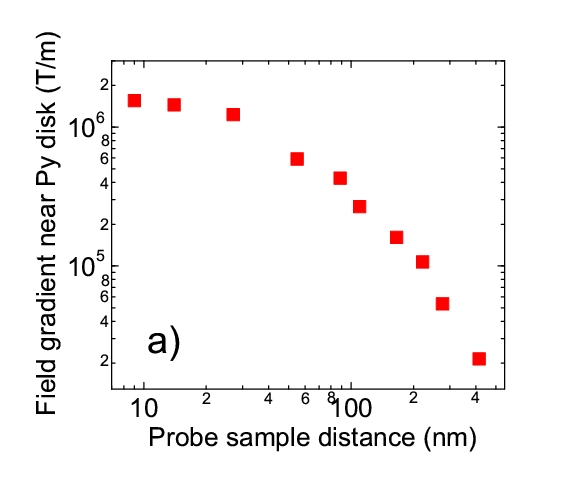}
\end{minipage}
\begin{minipage}[b]{4.2cm}
\vspace{0.2cm}
\includegraphics[width=4.6cm]{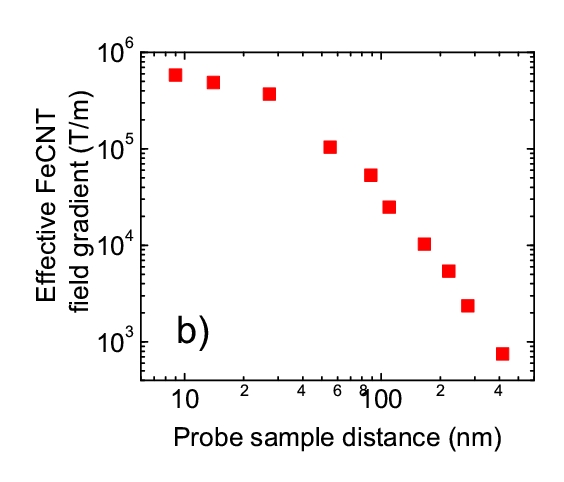}
\end{minipage}
\caption{a) Magnetic field gradient at point B (Fig.~\ref{fig:MFMvsZ}a) near the Py dot as a function of probe-sample separation using probe J2. b) The magnetic field gradient close to the tip of J2. Both graphs are obtained from the data in Fig.~\ref{fig:FreqVsZ} using the monopole model. $d_s$ denotes the separation between the sample surface and tip of the carbon nanotube; the iron wire end is located $\sim 30$\,nm further away.}
\label{fig:FeCNTandPy}
\end{figure}

These results demonstrate that a FeCNT is well modeled as a monopole with a magnetic charge positioned a short distance ($\sim 30$\,nm) from the CNT's end. This makes the FeCNT probe unique for quantitative MFM experiments. With knowledge of the monopole charge $Q$ we can directly image the magnetic field gradient generated by the sample:
 \begin{equation}
 \delta f(x,y,z) =
 \frac{f_0}{2k} Q \cdot \frac{\partial \bf H}{\partial \bf l}(x,y,z)
 \label{eg:SampleField}
 \end{equation}
In contrast to Eq.~\ref{eq:ForceGradient}, here $\frac{\partial \bf H}{\partial \bf l}$ is the sample field gradient. Using Eq.~\ref{eg:SampleField} we can extract the value of the sample's magnetic field gradient for various $d_s$ from the frequency shift data in Fig.~\ref{fig:FreqVsZ}. This evaluation is shown in Fig.~\ref{fig:FeCNTandPy}a for FeCNT probe J2. The measured probe sample separation is displayed without adding the carbon shell thickness of $\sim 30$\,nm. For $d_s \sim 20$\,nm we measured a field gradient of $1.5 \cdot 10 ^6$\,T/m. This is a relatively large gradient given that the saturation magnetization for permalloy is only $4 \pi M^{\rm Py}_s \sim 10^4$ G.

Using the proposed monopole model we can also reconstruct the field gradient near the tip of the FeCNT probe. The probe gradient is obtained by calculating $\frac{\partial H}{\partial z} = 2Q/(d_s+ d_w)^3$  ($d_w$ is the carbon shell thickness) with the probe monopole moment $Q$ and $d_w$ obtained from the best fit between of the model to the data in Fig.~\ref{fig:FreqVsZ}. For each frequency shift value, the shift between simulated and measured curve is evaluated separately, so a slightly different $d_w$ is obtained for each point. In particular for small $d_s$, $d_w$ exceeds the mean 30\,nm. This reflects the increased influence of additional effects, e.g., electrostatics or mechanical vibrations, and this complicates the exact reconstruction of the FeCNT's field gradient. Therefore we call it an effective field gradient. The results are presented in Fig.~\ref{fig:FeCNTandPy}b. The carbon shell thickness $d_w =  30$\,nm is again not included in the probe sample distance. A maximum field gradient of $\sim 6 \cdot 10 ^5$\,T/m is detected near the FeCNT tip at a probe-sample distance of $\sim 20$\,nm. This value is smaller than the field gradient generated by the Py disk at the same distance. Because its diameter is small very large field gradients are generated by the iron nanowire, but this arbitrarily close approach is prevented by the carbon shell wall thickness which is comparable or larger than the iron wire diameter. Decreasing the carbon wall thickness by changing the FeCNT growth conditions or by electron beam induced oxidation of parts of the shell in water atmosphere (see e.g., Ref.~\onlinecite{Wolny2008}) could enable high gradient FeCNT probes.

The micromagnetic tip field gradient is a key parameter for high resolution magnetic force detection. For Magnetic Resonance Force Microscopy (MRFM) it determines the minimum magnetic moment that can be detected. The maximum value of the field gradient reported to date is $4.2 \cdot 10 ^6$\,T/m \cite{r:TobaccoMosaic}. A FeCNT probe is uniquely suited for quantitative measurement of the field gradient generated by a nanoscale magnetic probe  designed for high resolution MRFM. In addition, if the carbon shell thickness can be reduced, an FeCNT could also be employed as a probe for high resolution, high sensitivity MRFM.

\begin{figure}
\includegraphics[width=1.0\columnwidth]{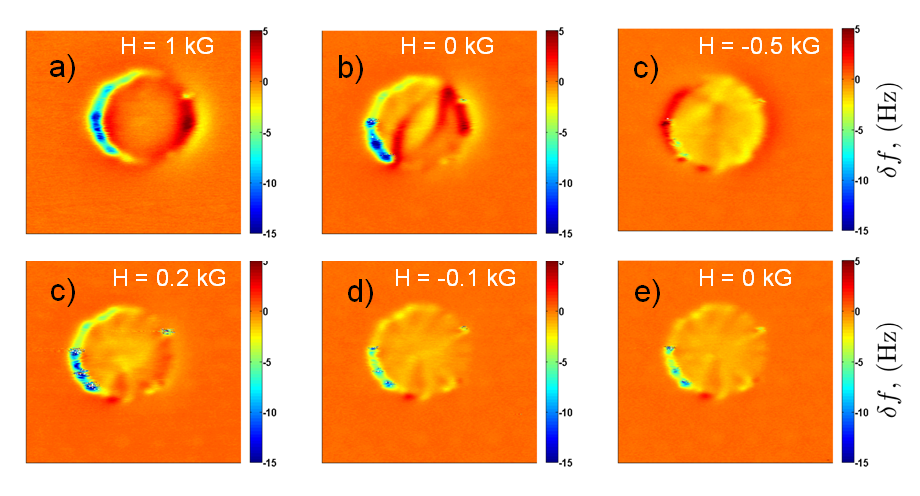}
\caption{MFM images (scan size 4.5\,$\mu$m\,$\times$\,4.5\,$\mu$m) of a Py disk made with the FeCNT probe J2 in different magnetic fields {\it perpendicular} to the sample plane. The magnetic field was applied in the following sequence: 2\,T (image not shown), 1\,kG, 0\,kG, -0.5\,kG, -0.2\,kG, -0.1\,kG, 0\,kG. The images were made at $T = 5$\,K.}
\label{fig:J2MFM}
\end{figure}

To demonstrate the applicability of an FeCNT as an MFM probe we performed MFM on Py disks in low magnetic fields. Fig.~\ref{fig:J2MFM} presents MFM images made with probe J2 in different external fields applied perpendicular to the sample plane with a $\sim 90$\,nm probe-sample separation and a $\sim 10$\,nm peak-to-peak cantilever oscillation amplitude at 5\,K. Due to the confirmed monopole character of the applied probe, these images reveal the local magnetic field gradient near the sample surface. Before the first measurement, a 2\,T was applied perpendicular to the sample; the field  was then slowly decreased. In an external field of 1\,kG the magnetic configuration is a single magnetic domain which evolves to a multidomain state near zero field. Here (Fig.~\ref{fig:J2MFM}b) the actual field on the sample is not exactly 0\,G, but a small remanent magnetic field from the external magnet remains.  The field was cycled with steadily decreasing amplitude to reduce this remanent field. At low fields, e.g., -0.5\,kG (Fig.~\ref{fig:J2MFM}c), we see a vortex-like magnetic structure centered slightly above the dot center, though the vortex core itself is not visible. In the next images of this sequence (Fig.~\ref{fig:J2MFM}d), the ``core" of the vortex shifts its $y$ (vertical) position possibly as a consequence of a non-zero tilt of the sample about the $\hat x$ axis.  Room temperature MFM measurements on the same material were presented in Ref.~\onlinecite{MFMvortices:T.Shinjo:Science2000}; there, the vortex core was clearly observed. This might be explained by the higher magnetocrystalline anisotropy of Py at low temperatures which would reduce the core size and make it less visible. This assumption is partially confirmed by measurements described below.

\begin{figure}
\includegraphics[width=1.0\columnwidth]{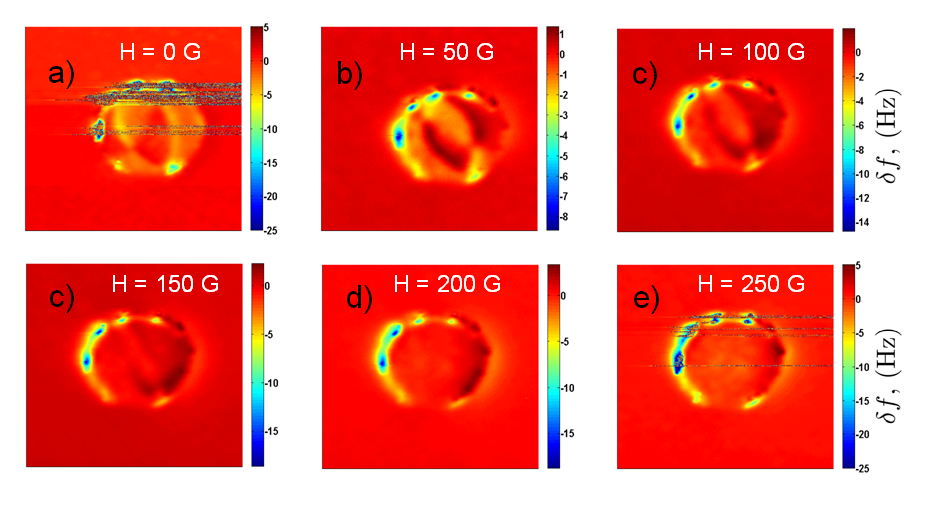}
\caption{MFM images (scan size 4.5 $\mu$m $\times$ 4.5 $\mu$m) of a Py disk made with the FeCNT probe K2 in different magnetic fields {\it parallel} to the sample plane. The magnetic field was applied in the following sequence: 0\,G, 50\,G, 100\,G, 150\,G, 200\,G, 250\,G. The images were made at $T = 30$\,K.}
\label{fig:K2MFM}
\end{figure}

Fig.~\ref{fig:K2MFM} shows MFM images of a Py disk at $T = 30$\,K and $d_s \sim 190$\,nm. The images were made with FeCNT probe K2 in varying external magnetic fields applied {\it parallel} to the sample plane. We have previously shown
that, due to their large shape anisotropy, FeCNT probes are suitable for measurements of in-plane fields \cite{Wolny2010}. Fig.~\ref{fig:K2MFM}a) taken in zero magnetic field, reveals domain structure distinct from a vortex state, and apparently similar to the three domain state observed in Fig.~\ref{fig:J2MFM}b).  With increasing external magnetic field the domain structure evolves and collapses at a field of 200\,G. For this particular Py disk the vortex structure could not be observed in any magnetic field or temperature, though it should be the minimum energy state. Multiple domains could result from pinning centers which prevent domain wall motion leading to a metastable domain structure.  Our observation of a vortex state in Fig.~\ref{fig:J2MFM}c) indicates each dot has a unique pinning structure.

In conclusion, we report a detailed characterization of two FeCNT MFM probes. We find that a FeCNT can be modeled as a monopole with a magnetic charge defined by the diameter of the iron wire enclosed in the carbon nanotube and the typical saturation magnetization of bcc iron. The carbon shell thickness near the iron wire edge was estimated to be $\sim 30$\,nm for FeCNT J2 and $\sim 60$\,nm for FeCNT K2. The FeCNT probes are uniquely suited for quantitative MFM. Knowing the magnetic charge of the monopole $Q$ we can directly image the field gradient generated by the sample. Quantitative measurements were performed on a Py disk, for which we measured a magnetic field gradient of $\sim 1.5 \cdot 10 ^6$\,T/m at a probe-sample distance of $\sim 20$\,nm. To demonstrate the use of the FeCNT as an MFM probe we imaged the magnetic structure of a Py disk in low external magnetic fields parallel and perpendicular to the sample plane. We find that the magnetic structure of the Py disks can be both vortex- and domain-like, probably depending on the individual properties of the Py disks.

This work was supported by the National Science Foundation through awards DMR-0807093 and the I2CAM Award, Grant DMR-0844115, by the Deutsche Forschungsgemeinschaft and by the ENCOMM Nanosystems Laboratory at OSU.


\begin{thebibliography}{10}

\bibitem{MagDomains}
A. Hubert and R. Schafer, {\em Magnetic Domains. The analysis of magnetic
  Microstructures} (Springer, Berlin, 1998), pp.\ 78--85.

\bibitem{Guntherodt:FirstMFM:jap:1987}
J.~J. Saenz, N. Garcia, P. Grutter, E. Meyer, H. Heinzemann, R. Wiesendanger,   L. Rosenthaler, H.~R. Hidber, and H.-J. Guntherodt, J. Appl. Phys. {\bf 62},
  4293  (1987).

\bibitem{Wiesendanger:magnetite:prb:2005}
M. Liebmann, A. Schwarz, U. Kaiser, R. Wiesendanger, D.-W. Kim, and T.-W. Noh,
  Phys. Rev. B {\bf 71},  104431  (2005).

\bibitem{Chang:MagPartice:jap:1993}
T. Chang, J.-G. Zhu, and J.~H. Judy, J. Appl. Phys. {\bf 73},  6716  (1993).

\bibitem{Hartmann:advances:2003}
M. Koblischka and U. Hartmann, Ultramicroscopy {\bf 97},  103  (2003).

\bibitem{Hug:QuantativeMFM:apl:1998}
H.~J. Hug, B. Stiefel, P.~J.~A. van Schendel, A. Moser, R. Hofer, S. Martin,
  H.-J. Gušntherodt, S. Porthun, L. Abelmann, J.~C. Lodder, G. Bochi, and R.~C. OHandley, J. Appl. Phys. {\bf 83},  5609  (1998).

\bibitem{Lohau:QuantitativeMFM:apl:2001}
J. Lohau, A. Carl, S. Kirsch, and E.~F. Wassermann, Appl. Phys. Lett. {\bf 78},   2020  (2001).

\bibitem{Hug:ExchBias:2005}
P. Kappenberger, I. Schmid, and H.~J. Hug, Advanced Engineering Materials {\bf
  7},  332  (2005).

\bibitem{Hug:TipModel:jap:1992}
A. Wadas and H.~J. Hug, J. Appl. Phys. {\bf 72},  203  (1992).

\bibitem{Hartman:Tip:1994}
G. Matteucci, M. Muccini, and U. Hartmann, Phys. Rev. B {\bf 50},  6823
  (1994).

\bibitem{Lohau:CoilMFM:jap:1999}
J. Lohau, S. Kirsch, A. Carl, G. Dumpich, and E.~F. Wassermann, J. Appl. Phys.
  {\bf 86},  3410  (1999).

\bibitem{Leonhardt2006}
A. Leonhardt, S. Hampel, C. M\"uller, I. M\"onch, R. Koseva, M. Ritschel, D.
  Elefant, K. Biedermann, and B. B\"uchner, Chem. Vap. Deposition {\bf 12},
  380  (2006).

\bibitem{Wolny2008}
F. Wolny, U. Weissker, T. M\"uhl, A. Leonhardt, S. Menzel, A. Winkler, and B.
  B\"uchner, Journal of Applied Physics {\bf 104},  064908  (2008).

\bibitem{Wolny2010}
F. Wolny, U. Weissker, T. M\"uhl, M.~U. Lutz, C. M\"uller, A. Leonhardt, and B.  B\"uchner, J. Phys.: Conf. Ser. {\bf 200},  112011  (2010).

\bibitem{winkler2006}
A. Winkler, T. M\"uhl, S. Menzel, R. Kozhuharova-Koseva, S. Hampel, A.
  Leonhardt, and B. B\"uchner, J. Appl. Phys. {\bf 99},  104905  (2006).

\bibitem{Mueller2006}
C. M\"uller, S. Hampel, D. Elefant, K. Biedermann, A. Leonhardt, M. Ritschel,
  and B. B\"uchner, Carbon {\bf 44},  1746  (2006).

\bibitem{sader:SpringConstant:64}
J.~E. Sader, Journal of Applied Physics {\bf 84},  64  (1998).

\bibitem{h:oboukhov.dsp}
Y. Obukhov, K.~C. Fong, D. Daughton, and P.~C. Hammel, J. Appl. Phys. {\bf
  101},  034315  (2007).

\bibitem{r:TobaccoMosaic}
C.~L. Degen, M. Poggio, H.~J. Mamin, C.~T. Rettner, and D. Rugar, Proceedings
  of the National Academy of Sciences {\bf 106},  1313  (2009).

\bibitem{MFMvortices:T.Shinjo:Science2000}
T. Shinjo, T. Okuno, R. Hassdorf, K. Shigeto, and T. Ono, Science {\bf 289},
  930  (2000).

\end{thebibliography}
\end{document}